\documentclass[11pt,twocolumn,prb,superscriptaddress,reprint,nohyperlinks]{revtex4-1}

\usepackage{ragged2e}
\usepackage{graphicx,amsmath,amssymb,color}
\usepackage{tabularx, ctable}
\usepackage{ulem}
\usepackage{bm}
\usepackage{siunitx}
\usepackage{gensymb}

\usepackage[breaklinks=true,colorlinks,allcolors=blue]{hyperref}

\def\re{\textcolor{black}}

\DeclareUnicodeCharacter{0301}{\'{e}}
\DeclareUnicodeCharacter{0301}{\'{0}}

\newcommand{\be}{\begin{eqnarray}}
\newcommand{\ee}{\end{eqnarray}}

\begin{document}

\title{Interaction-limited conductivity of twisted bilayer graphene revealed by giant terahertz photoresistance}

\author{A.L. Shilov$^{+}$}
\affiliation{Department of Materials Science and Engineering, National University of Singapore, 117575 Singapore}
\author{M. Kravtsov$^{+}$}
\affiliation{Department of Materials Science and Engineering, National University of Singapore, 117575 Singapore}
\affiliation{Institute for Functional Intelligent
Materials, National University of Singapore, Singapore, 117575, Singapore}

\author{J. Covey}
\affiliation{Department of Physics, University of Florida, Gainesville, FL 32611-8440, USA}

\author{M.A. Kashchenko}
\affiliation{Programmable Functional Materials Lab, Center for Neurophysics and Neuromorphic Technologies, 127495 Moscow}

\author{O. Popova}
\affiliation{Programmable Functional Materials Lab, Center for Neurophysics and Neuromorphic Technologies, 127495 Moscow}

\author{X. Zhou}
\affiliation{Department of Materials Science and Engineering, National University of Singapore, 117575 Singapore}

\author{L. Elesin}
\affiliation{Department of Materials Science and Engineering, National University of Singapore, 117575 Singapore}

\author{I.Yahniuk}
\affiliation{Terahertz Center, University of Regensburg, D-93053 Regensburg, Germany}

\author{T.~Taniguchi}
\affiliation{International Center for Materials Nanoarchitectonics, National Institute of Material Science, Tsukuba 305-0044, Japan}
\author{K. Watanabe}
\affiliation{Research Center for Functional Materials, National Institute of Material Science, Tsukuba 305-0044, Japan}

\author{A.I. Berdyugin}
\affiliation{Department of Materials Science and Engineering, National University of Singapore, 117575 Singapore}
\affiliation{Institute for Functional Intelligent Materials, National University of Singapore, Singapore, 117575, Singapore}

\author{Y. Wang}
\affiliation{Institute for Functional Intelligent Materials, National University of Singapore, Singapore, 117575, Singapore}

\author{S.D. Ganichev}
\affiliation{Terahertz Center, University of Regensburg, D-93053 Regensburg, Germany}
\affiliation{CENTERA Labs, Institute of High Pressure Physics, PAS, 01 - 142 Warsaw, Poland}

\author{V. Perebeinos}
\affiliation{Department of Electrical Engineering, University at Buffalo, The State University of New York, Buffalo, New York 14260, United States}

\author{D.A. Svintsov}
\affiliation{Moscow Center for Advanced Studies, Kulakova str. 20, Moscow, 123592, Russia}

\author{A. Principi}
\affiliation{School of Physics and Astronomy, University of Manchester, Manchester M13 9PL, United Kingdom}

\author{K.S. Novoselov}
\affiliation{Institute for Functional Intelligent Materials, National University of Singapore, Singapore, 117575, Singapore}

\author{D.L. Maslov}
\affiliation{Department of Physics, University of Florida, Gainesville, FL 32611-8440, USA}

\author{D.A. Bandurin}
\affiliation{Department of Materials Science and Engineering, National University of Singapore, 117575 Singapore}
\begin{abstract}


\textbf{Identifying the microscopic processes that limit conductivity is essential for understanding correlated and quantum-critical states in quantum materials. In twisted bilayer graphene (TBG) and other twist-controlled materials, the temperature ($T$) dependence of metallic resistivity ($\rho$) follows $\rho\sim T^\alpha$ scaling, with $\alpha$ spanning a broad range, rendering standard transport measurements insufficient to unambiguously identify the dominant scattering processes and giving rise to competing interpretations ranging from phonon-limited transport and umklapp scattering to strange metallicity and heavy fermion renormalization. 
Here, we use terahertz (THz) excitation to selectively raise the electron temperature in TBG while keeping the lattice cold, enabling a direct separation of electron–electron and electron–phonon contributions to resistivity. We observe a giant THz photoresistance -- reaching up to $70\%$ -- demonstrating that electronic interactions dominate transport even in regimes previously attributed to phonons, including the linear-in-$T$ resistivity near the magic angle. Away from the magic angle, we observe coexisting photoresistance and robust quadratic-in-$T$ resistivity at extremely low carrier densities where standard electron–electron scattering mechanisms (umklapp and Baber inter-band scattering) are kinematically forbidden. Our analysis identifies the breakdown of Galilean invariance in the Dirac-type dispersion \re{as a possible origin} of the interaction-limited conductivity, arising from \re{inter-valley electron-electron (\textit{e-e}) collisions}. Beyond twisted bilayer graphene, our approach establishes THz-driven hot-electron transport as a general framework for disentangling scattering mechanisms in low-density quantum materials.}

 \begin{center}
 \end{center}

\end{abstract}

\maketitle

Electrical resistivity is a key probe of the microscopic interactions that govern charge transport in metals. A paradigmatic example is the $T^2$ temperature dependence of resistivity ($\rho$), long recognized as a hallmark of electron–electron (\textit{e–e}) scattering in metals.\cite{Landau:1936idb,baber1937resistance}
The ubiquity of this scaling in both elemental \cite{PhysRevLett.20.1439} and heavy-fermion compounds \cite{KADOWAKI1986507} underpinned the development of Fermi liquid (FL) theory, where the $T^2$ dependence follows naturally from the quadratic growth of \textit{e-e} scattering within a thermal window of the Fermi level. Despite its foundational importance, the microscopic mechanisms of charge transport in materials hosting FLs have, beyond basic textbook descriptions, remained largely unaddressed. These mechanisms are inherently nontrivial because, in FLs, normal \textit{e-e} collisions conserve net charge current and do not degrade electrical conductivity. A finite $T^2$ resistivity thus requires a current-relaxing mechanism of electron-electron interaction. \cite{pal2012resistivity} Two established microscopic scenarios - umklapp and Baber inter-band scattering \cite{baber1937resistance} - were long believed sufficient to account for current dissipation in all FL metals. The universality of this paradigm  has been called into question by recent studies of $\rho(T)$ behaviour in dilute metallic systems, such as SrTiO$_3$\cite{doi:10.1126/science.aaa8655}, Bi$_2$O$_2$Se\cite{wang_t-square_2020}, and
HgTe quantum well \cite{PhysRevLett.134.196303}. These materials exhibit a robust $T^2$ resistivity even at vanishingly low electron doping levels, where the Fermi surface is reduced to a small, single-component pocket, and both umklapp and Baber mechanisms become inoperative\cite{doi:10.1126/science.aaa8655,wang_t-square_2020}. This reveals
profound gaps in understanding of how \textit{e-e} scattering in some FLs translates microscopically into $T^2$ resistivity behaviour \cite{https://doi.org/10.1002/andp.202100588}.

\begin{figure*}[ht!]
  \centering\includegraphics[width=0.95\linewidth]{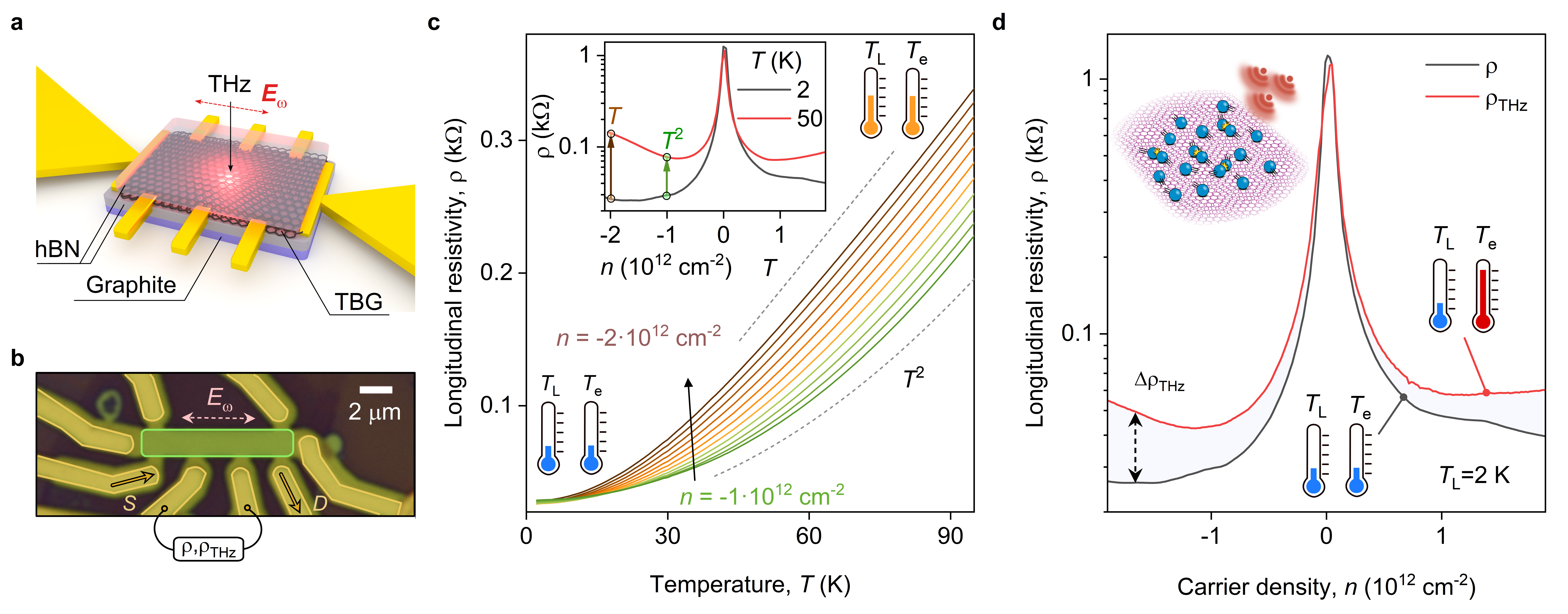}
    \caption{\textbf{Resistivity in TBG under THz excitation.} \textbf{a,} Cartoon schematic of a typical TBG device, showing THz-induced electron heating in the device’s channel. THz radiation funneling into the device is polarized along the channel. \textbf{b,} Optical image of a typical TBG device. Metallic contacts are highlighted in yellow, and the TBG channel exhibiting a THz-induced change in resistivity is highlighted in green. \textbf{c,} Longitudinal resistivity $\rho$ of $2\degree$ TBG as a function of temperature and carrier density $n$. Inset: $\rho(n)$ at representative temperatures. Arrows indicate scaling regimes with $\rho \sim T^\alpha$, where $\alpha = 1$ (orange) and $\alpha = 2$ (green). Thermometers indicate that both electron and lattice temperatures are varied simultaneously. \textbf{d,} $\rho(n)$ measured in the dark and under continuous-wave illumination at 0.14~THz. Inset: schematic showing THz-driven electron heating. Thermometers indicate that THz radiation increases the electron temperature while the lattice remains intact.}
    \label{fig:F1}
\end{figure*}


Concomitantly with the expansion of the FL family, a growing number of novel material platforms emerged, displaying unusual correlated electronic effects that depart from the conventional FL model. Complex microscopic mechanisms of electronic interactions in such non-Fermi liquid (NFL) phases give rise to a rich variety of $\rho(T)$ forms, notably diverging from the canonical $T^2$ dependence. Among NFL candidates, small-angle twisted bilayer graphene (TBG) stands out as a particularly versatile experimental platform that uniquely combines tunable band structure and multifaceted transport behaviour, suggestive of both FL and NFL physics. Transport measurements in TBG revealed diverse functional $\rho \sim T^\alpha$ dependencies, with scaling exponent $\alpha$ varying smoothly between 1 and 2 across the moir\'e band filling~\cite{PhysRevB.98.035425,jaoui_quantum_2022}.  Deviations from the typical FL picture in TBG become especially pronounced when the twist angle $\theta$ is tuned to the magic angle ($\approx 1.1\degree$), where electronic correlations are strongly enhanced due to the formation of flat moiré bands\cite{cao_unconventional_2018,cao_correlated_2018}. In proximity to correlated superconducting states near half-band filling, magic-angle TBG (MATBG) exhibits anomalously steep $T$-linear resistivity often associated with quantum criticality and strange metallicity\cite{PhysRevLett.124.076801,jaoui_quantum_2022,ghawri_breakdown_2022,PhysRevLett.127.266601}. Alternative interpretations invoke electron-phonon (\textit{e-ph})\cite{polshyn_large_2019,PhysRevB.99.140302,PhysRevB.94.245403,sharma_carrier_2021,PhysRevB.98.035425,DASSARMA2020168193,ishizuka_purcell-like_2021,PhysRevB.109.195105,PhysRevResearch.4.033061,sharma_carrier_2021,PhysRevB.99.165112,PhysRevB.94.245403} or \textit{e-e} umklapp \cite{Ishizuka_2022} scattering processes amplified by the flat band structure. Simultaneously, robust $T^2$ resistivity regime\cite{PhysRevB.98.035425} and characteristic quadratic-in-$T$ scattering lengths\cite{doi:10.1126/sciadv.aay7838} emerge at low carrier densities in MATBG and larger twist angles, suggesting a recovery of metallic FL behavior~\cite{jaoui_quantum_2022}. 
Quadratic resistivity has also been observed in magic-angle twisted trilayer graphene, which they attributed to heavy-fermion renormalization~\cite{Shuigang}.
Ultimately, whether TBG’s unusual resistivity trends originate from genuine non-FL phenomena, reflect a strongly renormalized FL, or arise from $\rho \!\sim\! T^{\alpha}$ scaling with variable $\alpha$ in an electron--phonon scattering scenario remains obscure.
The rich theoretical landscape \cite{PhysRevB.109.205102,PhysRevB.108.075168,PhysRevLett.124.186801,nano11051306,PhysRevB.100.155426} surrounding $\rho\sim T^\alpha$ functional forms in TBG reflects the ambiguity in separating contributions of electronic and phonon-mediated scattering mechanisms. Despite comprehensive transport characterization, analysis of the $\rho(T)$ scaling alone has thus far proven insufficient to definitively resolve the microscopic origin of conductivity in TBG.


In this work, we probe the resistivity of TBG under terahertz (THz) irradiation, which selectively heats the electronic system while leaving the lattice cold. This technique allows us to diagnose interaction-limited transport by isolating the contribution of electron--electron scattering from that of phonons. In response to the THz-driven increase in electron temperature $T_\mathrm{e}$, TBG exhibits a giant photoresistance, even in regimes where equilibrium resistivity is linear in $T$, indicating interaction-limited behaviour. Away from the magic angle, robust positive photoresistance coexists with $T^2$ resistivity---a hallmark of FL transport. Strikingly, the onset of $T^2$ scaling occurs at vanishingly small carrier densities where standard \textit{e-e} scattering mechanisms (umklapp and Baber inter-band scattering) are kinematically forbidden. Our  analysis identifies the breakdown of Galilean invariance in the Dirac-type dispersion of TBG \re{as a possible origin of the interaction-limited conductivity caused by intervalley \textit{e-e} collisions}.

\begin{figure*}[ht!]
  \centering\includegraphics[width=0.93\linewidth]{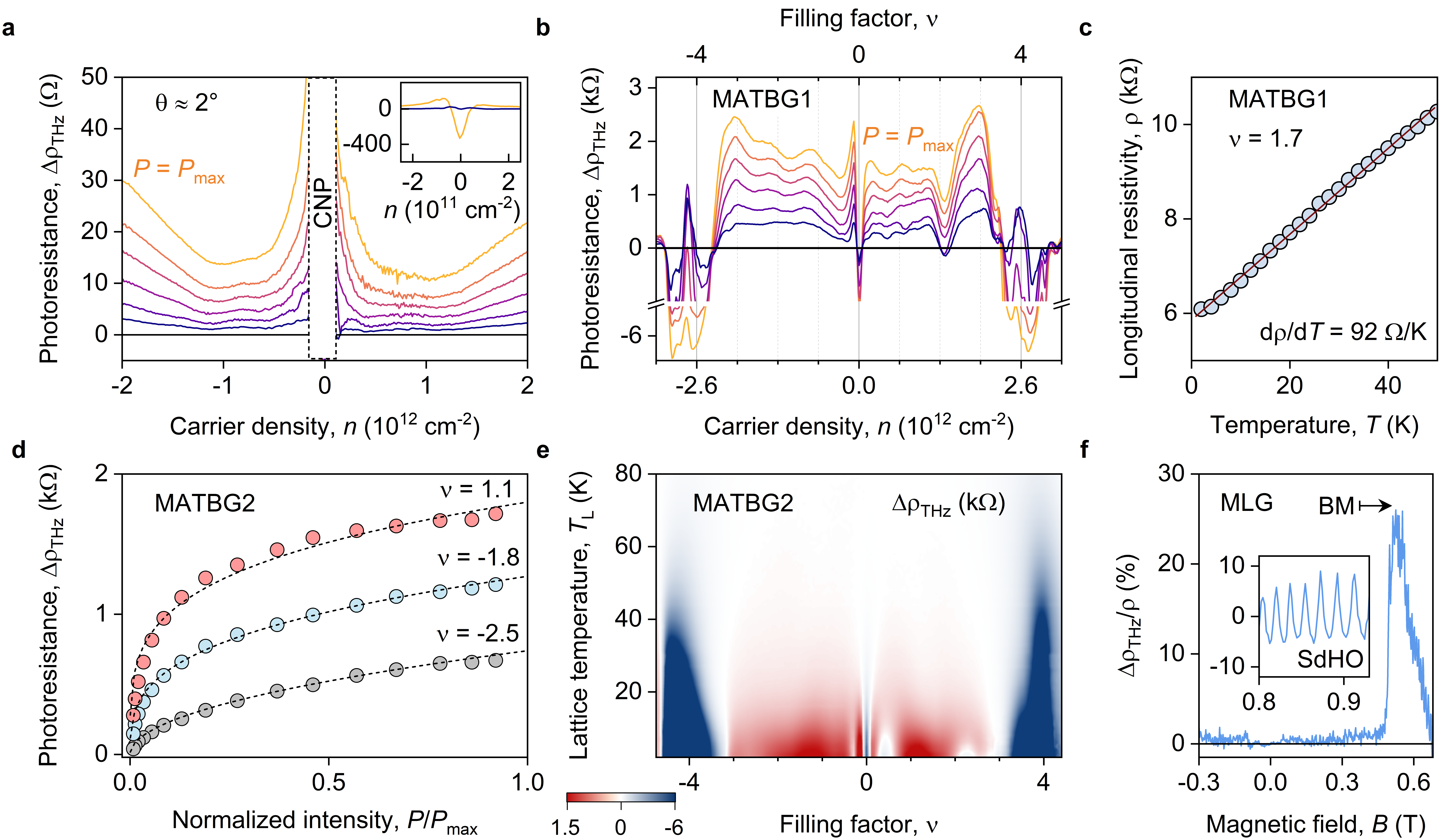}
    \caption{\textbf{Interaction-driven photoresistance in TBG.} \textbf{a,} $\Delta \rho_\mathrm{THz}$ as a function of $n$, plotted for radiation powers $P$ = 0.03 (blue), 0.09, 0.19, 0.37, 0.57, 1 (orange) in units of $P_\mathrm{max}$, for $2\degree$ TBG. Inset: $\Delta \rho_\mathrm{THz}$ close to the CNP for zero and maximum $P$. \textbf{b,} Same as (a) but for MATBG1 device ($\theta = 1.05\degree$). \textbf{c,} Longitudinal resistivity $\rho$ as a function of temperature $T$ for MATBG1 device at filling factor $\nu = 1.7$. The slope $d\rho/dT = 92~\mathrm{\Omega/K}$ is consistent with the Planckian bound in MATBG\cite{PhysRevLett.124.076801}. \textbf{d,} $\Delta \rho_\mathrm{THz}$ as a function of $P$ for given $n$ measured in MATBG2 device ($\theta = 1.04\degree$). Dashed lines - guides to the eye. \textbf{e,} $\Delta\rho_\mathrm{THz}$, plotted against $\nu$ at $T_\mathrm{L}$ for MATBG2.  \textbf{f,} $\Delta \rho_\mathrm{THz}$ in monolayer graphene, normalized to the dark resistivity $\rho$, as a function of magnetic field $B$. Inset: Zoomed-in view highlighting a finite $\Delta \rho_\mathrm{THz}$ response arising from $T_\mathrm{e}$-sensitive Shubnikov–de Haas oscillations (SdHO). Arrow labeled "BM" marks the position of the Bernstein mode resonance.
}
    \label{fig:F2}
\end{figure*}

\begin{figure*}[ht!]
  \centering\includegraphics[width=0.85\linewidth]{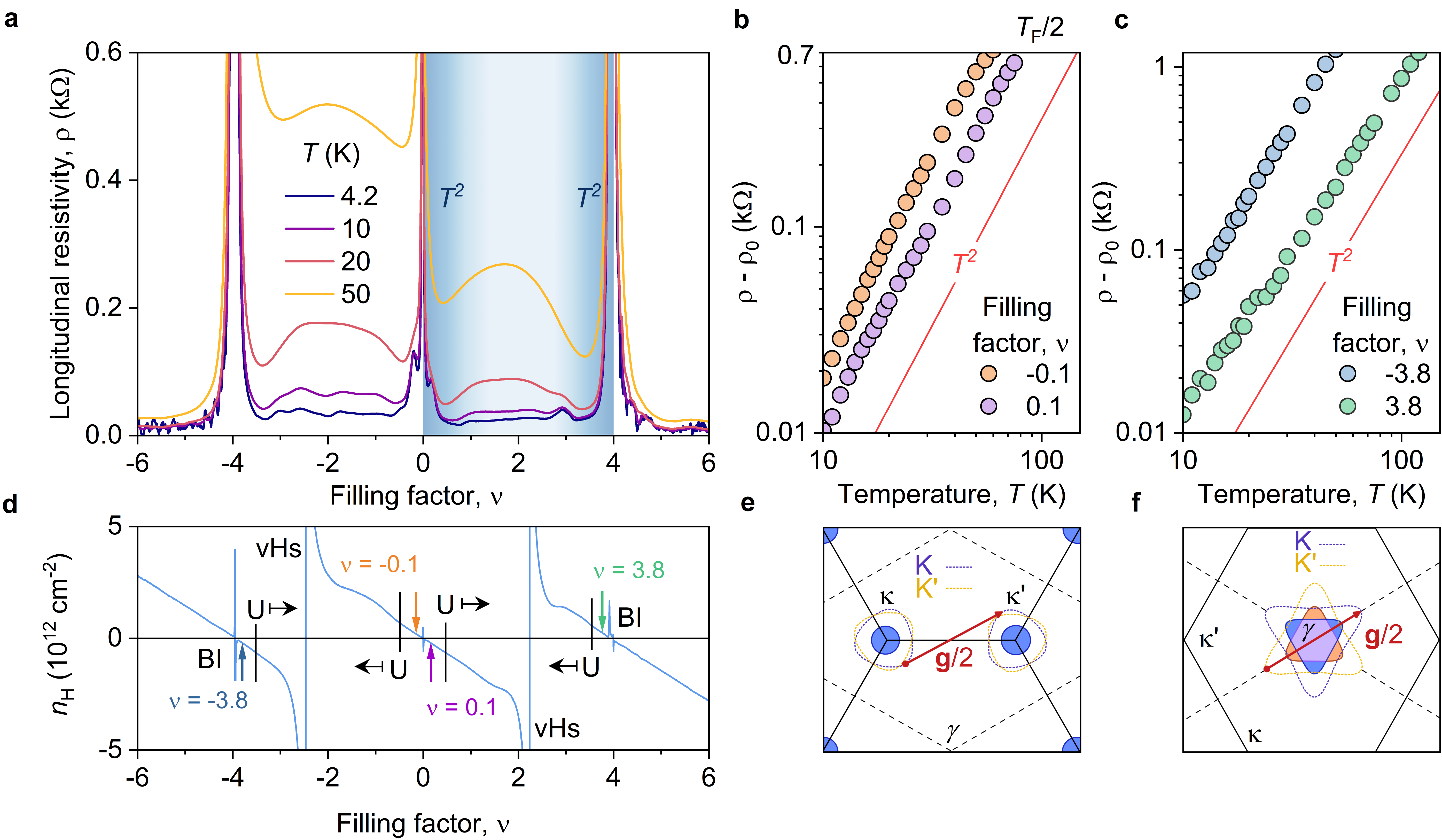}
\caption{\textbf{Anomalous $T^2$ resistivity in TBG.} 
\textbf{a,} Longitudinal resistivity $\rho$ as a function of filling factor $\nu$ at characteristic temperatures in $1.58\degree$ TBG. Gradient shading highlights a smooth crossover from $T^2$ to linear-$T$ and back to $T^2$ temperature dependence across different $\nu$. 
\textbf{b-c,} Temperature-dependent change in resistivity, $\Delta\rho = \rho - \rho_0$, shown for representative $\nu$ values near the CNP (b) and superlattice gaps (c).
\textbf{d,} Hall carrier density $n_\mathrm{H}$ versus $\nu$ for the same device. Arrows labeled “U” indicate the filling ranges within which intervalley umklapp scattering is allowed. Blue and orange arrows mark the onset of $T^2$ scaling in the hole-side miniband. Purple and green arrows mark the onset of $T^2$ scaling in the electron-side miniband.
\textbf{e,} Fermi surfaces at $\nu = 0.1$ (solid circles), where $\rho\sim T^2$ onsets, and $\nu = 0.45$ (dashed circles), where intervalley umklapp is allowed, plotted within the $1.58\degree$ mini Brillouin zone with high-symmetry points $\kappa$, $\kappa'$, and $\gamma$. Blue and orange colours indicate Fermi surfaces of the bands associated with K and K' symmetry points of the original BZ.
\textbf{f,} Same as (e), but for fillings in the vicinity of the electron-side superlattice gap. Solid and dashed lines mark the Fermi surfaces at $\nu = 3.8$ and $\nu = 3.45$, respectively.}
    \label{fig:F3}
\end{figure*}

 Figure \ref{fig:F1}a shows exemplary transport data that we measured in a typical TBG device (see Methods). Notably, the intriguing crossover between quasi-linear and quadratic $T$-scalings, traced as a function of band filling, persists even in bilayers with $\theta$ far from the magic angle, where the role of flat band correlations is allegedly diminished. Conflation of scattering mechanisms that are governed individually by electronic $T_\mathrm{e}$ and lattice $T_\mathrm{L}$ temperatures significantly impairs conventional $\rho(T)$ analysis. To circumvent these limitations, we exposed our devices to high-frequency radiation and measured the induced change of resistance (Fig.~\ref{fig:F1}b). 
The idea is that while in most materials $T_\mathrm{e}$ is tightly coupled to $T_\mathrm{L}$, in graphene-based systems electrons can become thermally detached from the lattice as a result of inherently weak electron-phonon coupling even in the case of small-angle ($<2\degree$) TBG where the latter is somewhat enhanced \cite{doi:10.1126/sciadv.adj1361}. 
Furthermore, TBG exhibits a relatively small electronic specific heat, such that even minimal power input rapidly drives the electron system out of equilibrium with the lattice~\cite{Battista}. A new hot carrier distribution is established on femtosecond timescales via carrier–carrier scattering, enabling us to probe effects that are exclusively sensitive to the $T_\mathrm{e}$ variation.
When exposed to a continuous-wave (CW) $f = 0.14$~THz beam, $\rho$ exhibits a pronounced increase at all carrier densities $n$ away from the CNP (Fig. \ref{fig:F1}b). 
Since 0.14~THz photons (0.6~meV) -- unlike infrared light\cite{Hubmann_2023} -- cannot induce interband transitions in doped TBG even at the magic angle, contributions from interband photoconductivity can be reliably dismissed. 
This ensures that the observed enhancement in $\rho$ originates solely from elevated $T_\mathrm{e}$.

To get deeper insights into the implications of THz-driven rise in $T_\mathrm{e}$ on TBG's transport, we measured photoresistance $\Delta \rho_\mathrm{THz} = \rho_\mathrm{THz}-\rho$ (here $\rho_\mathrm{THz}$ is the resistivity under exposure to THz radiation) for bilayers with different twist angles (Methods).
Figures \ref{fig:F2}a,b display examples of $\Delta \rho_\mathrm{THz}(n)$ for $\theta \approx 2 \degree$ and $\theta$ close to the magic-angle. For both $\theta$, THz-induced heating smears out the electronic distribution function, leading to a negative $\Delta\rho_\mathrm{THz}$ at the CNP. Similarly negative $\Delta \rho_\mathrm{THz}$ is observed at the full filling of the MATBG flat band $\nu=\pm4$ (here $\nu$ is the filling factor) where the single-particle band insulator emerges and at $\nu\approx2$ pointing to the presence of correlated  gap. At all $n$ away from these regions, $\Delta \rho_\mathrm{THz}$ is strong, positive and grows upon increasing THz power, $P$, for all studied $\theta$. This makes $\Delta \rho_\mathrm{THz}$ a convenient probe for distinguishing insulator-like and metallic-like responses to variations in $T_\mathrm{e}$ in TBG under cold lattice conditions.

The observed positive $\Delta \rho_\mathrm{THz}$ in doped TBG away from $\nu = 0, 2, \pm4$ indicates a strong contribution of electron interactions to its resistivity. This conclusion follows from the fact that  while $T_\mathrm{L} = 2$~K is held constant, $\rho$ grows substantially with rising $T_\mathrm{e}$. In our THz experiments,  $\Delta\rho_\mathrm{THz}$ increases significantly with rising $P$ and saturates towards $P=P_\mathrm{max}$ (Fig.~\ref{fig:F2}d), akin to the photoconductivity observed in TBG at marginally small $\theta$ \cite{PhysRevMaterials.6.024003}. From the comparison of $\Delta\rho_\mathrm{THz}$ with standard transport $\rho(T)$ measurements and the analysis of the THz-induced damping of Shubnikov–de Haas oscillations (SdHO), which are sensitive to the smearing of the electronic distribution function, we estimate that electrons can be thermalized at $T_\mathrm{e}$ up to 15~K above the cold lattice at the highest $P$ (Supplementary Information). As expected, $\Delta \rho_\mathrm{THz}$ vanishes with increasing $T_\mathrm{L}$, due to more efficient electron cooling (Fig.~\ref{fig:F2}e). 


In magic-angle devices, $\Delta \rho_\mathrm{THz}$ is exceptionally large, with $\Delta \rho_\mathrm{THz}/\rho$ reaching up to 70\% (see Supplementary Information). Remarkably, $\Delta \rho_\mathrm{THz}$ remains positive even at carrier densities where $\rho \sim T$ (Fig.~\ref{fig:F2}c). In devices with larger twist angles, $\Delta \rho_\mathrm{THz}$ is also strong and positive, regardless of whether $\rho$ scales linearly or quadratically with $T$. These observations indicate that \textit{e-e} interactions drive the temperature-dependent rise in resistivity, even in regimes where $\rho$ is linear in $T$, that is often misattributed to \textit{e-ph} scattering\cite{polshyn_large_2019,PhysRevB.99.140302,PhysRevB.94.245403,sharma_carrier_2021,PhysRevB.98.035425,DASSARMA2020168193,ishizuka_purcell-like_2021,PhysRevB.109.195105,PhysRevResearch.4.033061,sharma_carrier_2021,PhysRevB.99.165112,PhysRevB.94.245403}.
Indeed, at $T_\mathrm{L} = 2$~K, \textit{e-ph} scattering can, in principle, result in a finite $\Delta\rho_\mathrm{THz}$ via acoustic phonon emission by hot electrons, but its contribution is negligible in our experiments. This is  evident from the data obtained on a sample made of monolayer graphene (MLG) featuring phonon-limited conductivity. When doped, the device exhibits vanishingly small $\Delta \rho_\mathrm{THz}$, with a measurable response emerging only upon application of a finite magnetic field $B$, due to $T_\mathrm{e}$-sensitive effects such as Bernstein mode (BM) resonances or SdHO (Fig.~\ref{fig:F2}f)~\cite{Bandurin2022,TIMO}.

Having established that electron interactions have strong contributions to the $T-$dependent resistivity of TBG, we next examine possible microscopic mechanisms underlying current relaxation. To this end, we focus more closely on $\rho$ in the device with a twist angle $\theta = 1.58\degree$ (Fig.~\ref{fig:F3}). At this $\theta$, the TBG Fermi velocity is substantially reduced relative to MLG, yet not low enough to observe correlated insulator states and superconductivity. Importantly, the low disorder in this device enabled metallic conductivity down to vanishingly small doping levels, which allowed us to probe functional forms of  $\rho(T)$ at carrier densities as low as $10^{11}$~cm$^{-2}$. 

Figure~\ref{fig:F3}a shows examples of $\rho(\nu)$ dependencies measured on this device at various $T$ and reveals $\rho \sim T^2$ regimes close to the neutrality point and band insulator (blue shaded region in Fig~\ref{fig:F3}a). The presence of $\rho \sim T^2$ regimes at both low and high doping naturally prompts the question of whether \textit{e–e} umklapp scattering is the dominant mechanism limiting TBG conductivity at elevated temperatures, as proposed in theory~\cite{Ishizuka_2022}. 
Indeed, the moiré potential dramatically shrinks the Brillouin zone (BZ), allowing electrons to scatter between adjacent minizones and thereby enabling strong backscattering and resistance.  
Specifically, umklapp process requires that the total momentum change $\Delta k$ of a colliding electron pair equals a reciprocal lattice vector g. For a hexagonal BZ with Fermi pockets located at the $\kappa$ and $\kappa'$ points, this imposes a threshold condition on the Fermi momentum $k_\mathrm{F} > g/4\sqrt{3}$ (Fig.~\ref{fig:F3}e). Likewise, when the first miniband is nearly completely filled and Fermi pockets reside around the $\gamma$ point of the BZ, the umklapp threshold becomes $k_\mathrm{F} > g/4$ (Fig.~\ref{fig:F3}f). For a $\theta = 1.58\degree$ device, these conditions define doping ranges within a first pair of minibands, where umklapp scattering is allowed --  namely, $-3.5<\nu<0.5$ in a hole-side band and $0.45<\nu<3.45$ in an electron-side band. Indeed, analysis of the low-$T$ range of resistivity in the 1.58\degree~device reveals a clear enhancement of the $T^2$ prefactor within these fillings (see Supplementary Information). However $T^2$ growth remains appreciable even at fillings as low as $\nu \sim 0.1$ (Fig.~\ref{fig:F3}b,c), well below the momentum threshold for umklapp processes, suggesting that additional scattering mechanisms, governed by $T_\mathrm{e}$, are involved in current relaxation. Robust $T^2$ resistivity in umklapp-free regime is consistently reproduced in all our high-angle ($\theta = 1.6-3\degree$) devices, and has also been previosuly reported for MATBG proximitized by a screening metallic layer\cite{jaoui_quantum_2022}. We also point out a striking resemblance between this anomalous resistivity behaviour in TBG and the mysterious $T^2$ scaling in SrTiO$_3$\cite{doi:10.1126/science.aaa8655} and Bi$_2$O$_2$Se\cite{wang_t-square_2020}, which likewise persists down to marginally small densities. While a $T^2$ resistivity in weakly doped SrTiO$_3$ can be tentatively attributed to two-phonon scattering \cite{kumar_STO:2021}, its origin in two other systems remains elusive.



We now turn to discuss other interaction-driven channels that can cause current dissipation. One example is electron--hole friction, which produces a $T^2$ contribution to resistivity in charge-neutral TBG under strong displacement fields that transforms it into a compensated semimetal with equivalent electron and hole Fermi pockets~\cite{Friction}. If the compensation is incomplete, electron--hole drag can arise, leading to an additional resistivity due to the entrainment of minority carriers against their expected motion in an electric field~\cite{Ponomarenko2024}. In the $n$ and $T$ range relevant to this study, both effects are strongly suppressed. At $|\nu| = 0.1$, where we observe the anomalous $T^2$ scaling, the Fermi energy is $\sim 20\ \mathrm{meV}$, ensuring that the electron system remains unipolar at all measured $T$.

If two most common channels -- umklapp and electron-hole scattering -- are ruled out at  $|\nu|=0.1$ and $3.8$, we need to consider other mechanisms that can be responsible for the observed $T^2$ scaling. To this end, we note that TBG electron system is characterized by the absence of Galilean invariance that is inherently broken due to the Dirac-type dispersion, and hence the momentum conservation due to normal \textit{e-e} collisions does not necessarily imply velocity and, therefore, current conservation. However, non-parabolic dispersion \textit{per se} does not guarantee a $T^2$ resistivity due to normal \textit{e-e} collisions; additional conditions are required, such as a singly connected concave Fermi surface or the presence of multiple non-equivalent valleys.
In TBG, trigonal warping lifts valley and layer equivalence (Fig.~\ref{fig:F3}e-f), enabling momentum-conserving \textit{e-e} scattering between distorted Fermi pockets. 
This ``interpocket'' scattering can generate a $T^2$ resistivity contribution~\cite{Sharma2021OpticalLiquid} that in the low-temperature limit is expressed through (See Methods): 
\begin{equation}
    \rho_\mathrm{ee} = \frac{\pi}{3}\frac{(k_\mathrm{F}a)^2}{ne^2v_\mathrm{F}^2}\alpha_e^2\frac{(k_\mathrm{B}T)^2}{\hbar}\Lambda, \label{rhoee}
\end{equation}
where $v_\mathrm{F}$ is the Fermi velocity, $\alpha_e = \frac{e^2}{4\pi\varepsilon_0\hbar v_\mathrm{F}}$ is the effective fine-structure constant, $a$ is the superlattice period, and $\Lambda$ is a dimensionless parameter capturing the microscopic nature of \textit{e-e} interactions in TBG. 
We note in passing that the $T^2$ growth does not persist indefinitely: theory~\cite{Gurzhi,Sharma2021OpticalLiquid} predicts its eventual saturation, beyond which electron–phonon scattering inevitably takes over.
Depending on the assumed screening model, $\Lambda$ ranges between $0.01$ and $1$, corresponding to a prefactor $A\simeq 0.005{-}0.5~\Omega/\mathrm{K}^2$ at $\theta=1.58^\circ$. Recent microscopic calculations of the effective interaction in TBG further confirm that this spread reflects genuine sensitivity to screening and Wannier form-factor effects~\cite{Sato2025}. Our measured prefactor of order $0.1~\Omega/\mathrm{K}^2$ falls within the theoretically expected range. Last, we note that steep resistivity growth in the first moiré band is succeeded by a relative weak $T-$dependence above $|\nu|=4$ fillings even though the effective mass and the Fermi surface size are comparable (Fig.~\ref{fig:F3}a). This indicates that it is the first (flat) moiré band properties, such as quantum geometry~\cite{Tian2023,VicQM} or Wannier localization~\cite{Sato2025}, that are responsible for the observed resistivity scaling. 




\section*{Conclusion}

We introduce a terahertz-driven hot-electron method to disentangle \textit{e-e} and \textit{e-ph} contributions to transport, applying it to TBG, where the microscopic origin of resistivity remains under debate. At the magic angle, we observe exceptionally large photoresistance---up to 70\%---demonstrating that electronic interactions dominate even in the linear-in-$T$ regime often attributed to phonons. At higher twist angles, robust $T^{2}$ resistivity persists below the umklapp threshold, indicating a current-relaxation mechanism enabled by the breakdown of Galilean invariance in Dirac bands via normal intervalley scattering. These findings unify the anomalous transport of magic-angle devices with the quadratic scaling of higher-angle systems, link TBG to other dilute correlated metals with interaction-limited conductivity, and establish a broadly applicable framework for understanding transport in moir\'e and other low-density quantum materials.

\section*{Methods}
\textbf{Device fabrication.}
Our samples were assembled from exfoliated hBN and graphite crystals using a standard dry-transfer technique described elsewhere\cite{Purdie2018}. To ensure moir\'e periodicity between the TBG layers, single-layer graphene films were mechanically torn in half and sequentially picked up at a small angle, precisely set by an automated rotation stage. The bilayers were encapsulated between two hBN slabs and equipped with graphite bottom gates to control carrier density. The samples were released onto the undoped insulating Si/SiO$_2$ substrates to minimize reflection of incident THz radiation. The Hall bar geometry of the devices and Ti/Au 1D metal contacts to TBG were defined using electron beam lithography followed by reactive ion etching in CHF$_3$/O$_2$ plasma. The source and drain electrodes were connected to the broadband triangular coupling the incident THz radiation with bilayers.

\textbf{Photoresistance measurements.} We measured our devices in a low-temperature (2-300 K) 7 T optical cryostat from Quantum Design. The setup comprised a $f = 0.14$~THz Terasense source and an optical path including a mirror and lenses that were used to direct the radiation beam to the sample (Supplementary Information). To obtain photoresistance $\Delta\rho_\mathrm{THz}$, we employed an approach based on dual modulation of excitation current $I$ ($f_\mathrm{I} = 17.17$~Hz) and output THz power $P$ ($f_\mathrm{mod} = 44.4$~Hz). To retrieve $\Delta\rho_\mathrm{THz}$ from the multi-harmonic voltage signal $V(t)$ built up across the device, the Lock-in amplifier was configured to measure Fourier component $V_\mathrm{dm}$ at $f_\mathrm{dm} = f_\mathrm{mod} - f_\mathrm{I}$, yielding a simple expression for $\Delta \rho_\mathrm{THz}$
\begin{equation*}
    \Delta \rho_\mathrm{THz} = \pi\frac{V_\mathrm{dm}}{I}\frac{d}{l},
\end{equation*}
where $l$ and $d$ are the length and width of the device, respectively, and dual modulation approach manifests in a constant pre-factor $\pi$. More detailed description of THz-driven photoresistance analysis can be found in our prior works.\cite{shilov_high-mobility_2024, kravtsov_viscous_2025}

\textbf{Derivation of interpocket $e-e$ scattering resistivity contribution.}
In our theoretical analysis, we follow the argument of Sharma \textit{et al.}\cite{Sharma2021OpticalLiquid}, developed for optical conductivity of a Dirac-Fermi liquid. In the dc-limit ($\omega\rightarrow0$), we obtain Eq.\eqref{rhoee}, with 
\begin{equation}
    \Lambda = \int\limits_0^2\frac{\overline{U}^2(x)}{x}\frac{\left(\Delta{\mathbf{v}}^\mathrm{TW}\right)^2(x)}{(v_\mathrm{F}k_\mathrm{F}a)^2}dx,\label{F}
\end{equation}
where $x = k/k_\mathrm{F}$ is a normalized momentum,
\begin{equation}
    \overline{U}(x) = \frac{2\varepsilon_0k_\mathrm{F}}{ e^2}U(xk_\mathrm{F})
\end{equation}
is the dimensionless $e-e$ interaction potential $U(k)$ in the momentum representation, and
\begin{equation}
\begin{split}
\left(\Delta{\mathbf{v}}^\mathrm{TW}\right)^2&(x) 
 =
\frac{(v_\mathrm{F}k_\mathrm{F}a)^2}{4}\\ \times\Bigg[13 (1 + x^2) &
- \frac{4 x^8 + 26 x^6 - 87 x^4 - 64 x^2 + 13}{(1+2x^2)^{3/2}} \Bigg]
\end{split}
\end{equation}
is the angle-averaged contribution from the trigonally warped part of TBG's dispersion to the squared change in velocity.

We consider two model scenarios of \textit{e-e} interactions. In the first case, Coulomb interactions are screened intrinsically, described within the Thomas–Fermi approximation by the effective potential characterized by the screening wave vector $k_\mathrm{TF} = 8\frac{e^2 E_\mathrm{F}}{4\pi\varepsilon_0 \hbar^2v_\mathrm{F}^2} = 8\alpha_e k_\mathrm{F}$. The corresponding interaction takes the form
\begin{equation}
    \overline{U}_\mathrm{TF}(x) = \frac{1}{x+k_\mathrm{TF}/k_\mathrm{F}}.
\end{equation}
Plugging this into Eq.(\ref{F}) and calculating the integral yields $\Lambda\approx0.01$. 

In the second model, we neglect the intrinsic screening of TBG and only account for electrostatic screening by the metallic gate, mediated by the dielectric hBN buffer separating it from TBG. In this case, the effective potential reads 
\begin{equation}
    \overline{U}(x)=\frac{1}{x\sqrt{\varepsilon_\perp\varepsilon_\parallel}}\left(1 - \mathrm{exp}(-\beta x)\right),
\end{equation}
where $\varepsilon_\parallel$ and $\varepsilon_\perp$ denote the in-plane and out-of-plane dielectric constants of hBN, respectively, and $\beta = 2k_\mathrm{F}d\sqrt{\varepsilon_\parallel/\varepsilon_\perp}$ is the geometric screening parameter with $d$ the hBN thickness. For the typical hBN thickness $d\simeq 50~$nm and $\varepsilon_\perp = 3.9$ and $\varepsilon_\parallel = 6.5$, we obtain $\Lambda \simeq 1$ at the relevant carrier densities.

In practice, the screening is expected to fall between the two limiting regimes, with $\Lambda$ taking values in the range $0.01$–$1$ (the best correspondence with experiment is achieved for $\Lambda \simeq 0.2$–$0.3$). It is important to note, however, that  the $T^2$ scaling form of the resistivity, arising from the interpocket \textit{e-e} scattering, is valid for any form of the interaction, as long as $U(k\rightarrow0) =$~const and $U(k\rightarrow\infty) =0$. Under these conditions, the microscopic details of the \textit{e–e} interaction affect only the magnitude of the $T^2$ prefactor.

\begin{flushleft}
\section*{Acknowledgments}
\justifying{This project is supported by MOE Tier 2 (Award T2EP50123-0020 given to D.A.B.). M.A.K. and  O.P. acknowledge the support of an internal funding program from the Center for Neurophysics and Neuromorphic Technologies. K.S.N. is grateful to the Ministry of Education, Singapore (Research Centre of Excellence award to the Institute for Functional Intelligent Materials, I-FIM, project No. EDUNC-33-18-279-V12) and to the Royal Society (UK, grant number RSRP\textbackslash R\textbackslash190000) for support. S.D.G. and I.Y. acknowledge the support of the Deutsche Forschungsgemeinschaft (DFG, German Research Foundation) via project WU 883/3-1.}
\end{flushleft}

\bibliography{Bibliography.bib}
\bibliographystyle{naturemag}

\newpage
\setcounter{figure}{0}
\renewcommand{\thesection}{}
\renewcommand{\thesubsection}{S\arabic{subsection}}
\renewcommand{\theequation} {S\arabic{equation}}
\renewcommand{\thefigure} {S\arabic{figure}}
\renewcommand{\thetable} {S\arabic{table}}
\end{document}